\documentclass[preprint,12pt]{elsarticle}%
\usepackage{amssymb}
\usepackage{amsfonts}
\usepackage{amsmath}
\usepackage{graphicx}%
\providecommand{\U}[1]{\protect\rule{.1in}{.1in}}
\journal{journal}

\begin{document}
\bigskip\bigskip%
\begin{frontmatter}%


%

\title
{A JKR solution for a ball-in-socket contact geometry as a bi-stable adhesive system}%

%

\author{M. Ciavarella}%
%

\address
{Politecnico di BARI. Center of Excellence in Computational Mechanics. Viale Gentile 182, 70126 Bari. Mciava@poliba.it}%
%

\begin{abstract}%

In the present note, we start by observing that in the classical JKR theory of
adhesion, using the usual Hertzian approximations, the pull-off load grows
unbounded when the clearance goes to zero in a conformal "ball in socket"
geometry. To consider the case of the conforming geometry, we use a recent
rigorous general extension of the original JKR energetic derivation proposed
by the first author which necessitates only of adhesionless solutions, and an
approximate adhesionless solution given in the literature. We find that
depending on a single governing parameter of the problem, $\theta=\Delta
R/\left(  2\pi wR/E^{\ast}\right)  $ where $E^{\ast}$ is the plane strain
elastic modulus of the material couple, $w$ the surface energy, $\Delta R$ the
clearance and $R$ the radius of the sphere, the system shows the classical
bistable behaviour for a single sinusoid or a dimpled surface: pull off is
approximately that of the JKR theory for $\theta>0.82$ only if the system is
not "pushed" strongly enough and otherwise a "strong adhesion" regime is
found. Below this value $\theta<0.82$, a strong spontaneous adhesion regime is
found similar to "full contact". From the strong regime, pull-off will require
a separate investigation depending on the actual system at hand. %

\end{abstract}%
%

%

\end{frontmatter}%



\section{\bigskip Introduction}

Adhesion has received very large attention recently, especially for insects
adhesion and bioinspired adhesives, the implementation of patterned surfaces
with pillars of various shapes and tips, and the understanding of mechanisms
of adhesion on nano- and micro-rough surfaces has greatly improved in recent
years. A typical assumption made is that of the assumption of JKR model
(Johnson et al., 1971) which corresponds to very short range adhesion where
adhesive forces are all within the contact area. For infinite surfaces, like
those composed of a single scale of sinusoids (Johnson, 1995), there is a
mechanism for bistable adhesion, also further simplified in the "dimple" model
of McMeeking et al. (2010), which essentially consists in a single depression
in one of the surfaces. 

Bi-stable adhesive systems, in which weak adhesion can be converted to strong
adhesion by the application of pressure, are very attractive because
switchable adhesion has certainly practical utility in the development of
advanced adhesion systems.

The geometry of a "ball-in-a-socket" has not received attention in the
adhesion community, despite "ball joints" are very common in industry as well
as in nature. In automobiles, spherical bearings connect control arms to the
steering knuckles. Ball-and-socket design is also clearly responsible of
functioning of the human hip joint, and the shoulder one, both of which may
receive prosthesis. Malfunctioning because of lack of lubrication results in
noise, failure, and adhesion of the joint (Brockett et al., 2007, Wang et al.,
2015). A sketch is represented in Fig.1.

\begin{center}%
\begin{tabular}
[c]{ll}%
{\includegraphics[
height=2.9415in,
width=2.8385in
]%
{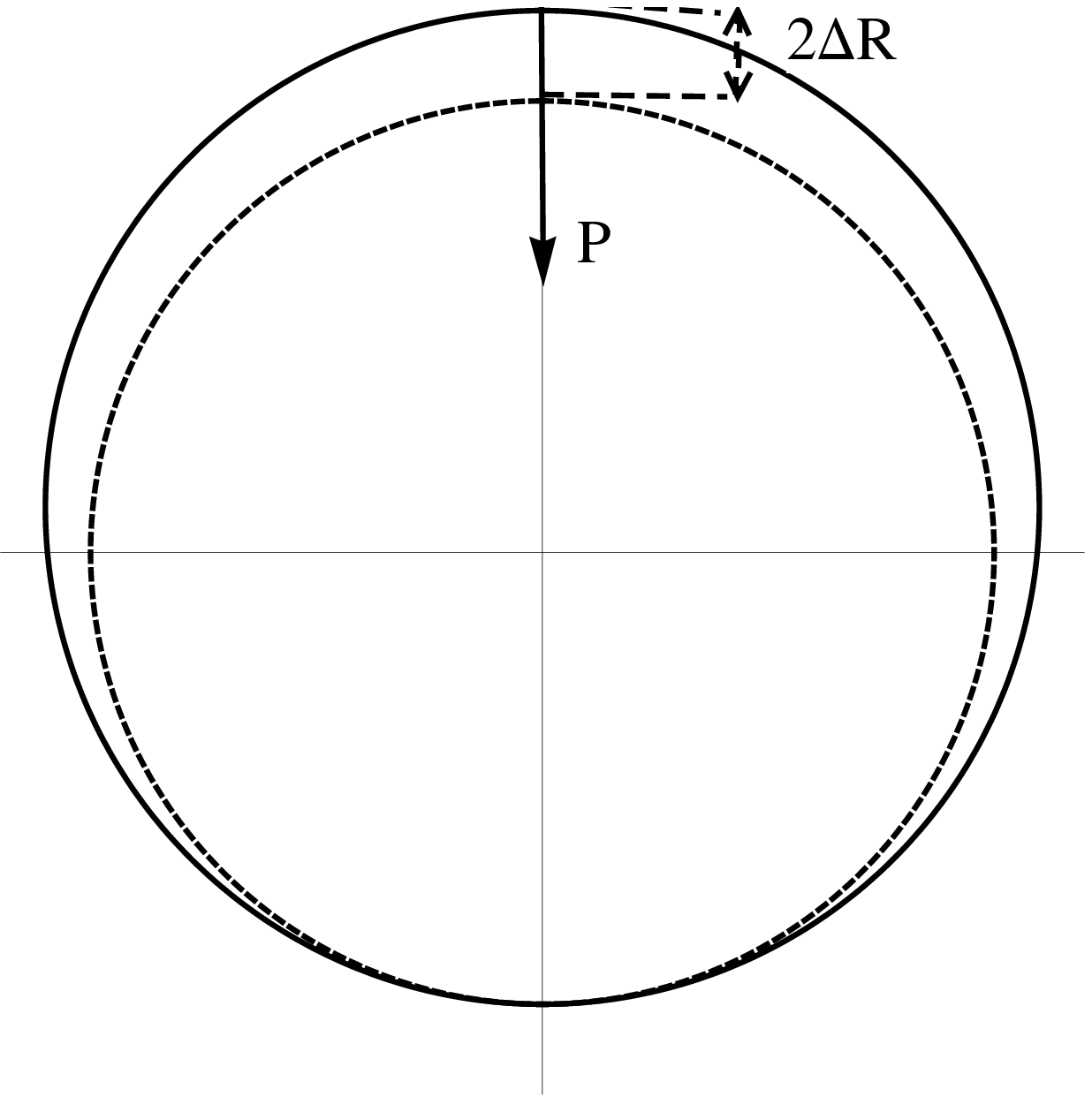}%
}
&
\end{tabular}

Fig.1. A sketch of a ball joint with clearance $\Delta R$
\end{center}

The ball joint can be analyzed for large clearances using the Hertz framework
(in the adhesionless conditions), or the JKR solution (Johnson et al., 1971),
with adhesion. In particular, the pull-off load for a ball-in-a-socket of
radius $R_{2}$ and of clearance $\Delta R$ is
\begin{equation}
P_{JKR}=-\frac{3}{2}\pi wR_{eq}=-\frac{3}{2}\pi w\frac{R_{2}^{2}}{\Delta R}%
\end{equation}
where $w$ is surface energy, $R=R_{2}$ is radius of the socket, and $\Delta R$
the clearance, where we recognize that the equivalent radius is that of
adhesionless Hertz' theory (Johnson, 1985)
\begin{equation}
\frac{1}{R_{eq}}=\frac{1}{R_{1}}-\frac{1}{R_{2}}=\frac{R_{2}-R_{1}}{R_{1}%
R_{2}}\simeq\frac{\Delta R}{R_{2}^{2}}%
\end{equation}

In other words, it is very important to remark that this pull-off value tends
to unbounded values for zero clearance a result which has not been remarked
before in the best of the author's knowledge, and calls for some attention.
Clearly, this stems from using Hertz theory rather than a correct conformal
theory. But we shall see that the situation is actually more complex. Solving
the JKR problem using the full elasticity of the ball-in-socket problem is the
scope of the present paper. The problem would seem very complicated, but in a
recent note (Ciavarella, 2017), we suggested a simple closed form solution to
the adhesive contact problem under the so-called JKR regime, which could be
exact under special symmetry conditions, including the present geometry,
whereas it was presented in that note as an approximate general result. The
derivation is based on generalizing the original JKR energetic derivation
assuming calculation of the strain energy in adhesiveless contact, and
unloading at constant contact area (see Fig.2).

\begin{center}%
\begin{tabular}
[c]{ll}%
{\includegraphics[
height=3.3728in,
width=5.0571in
]%
{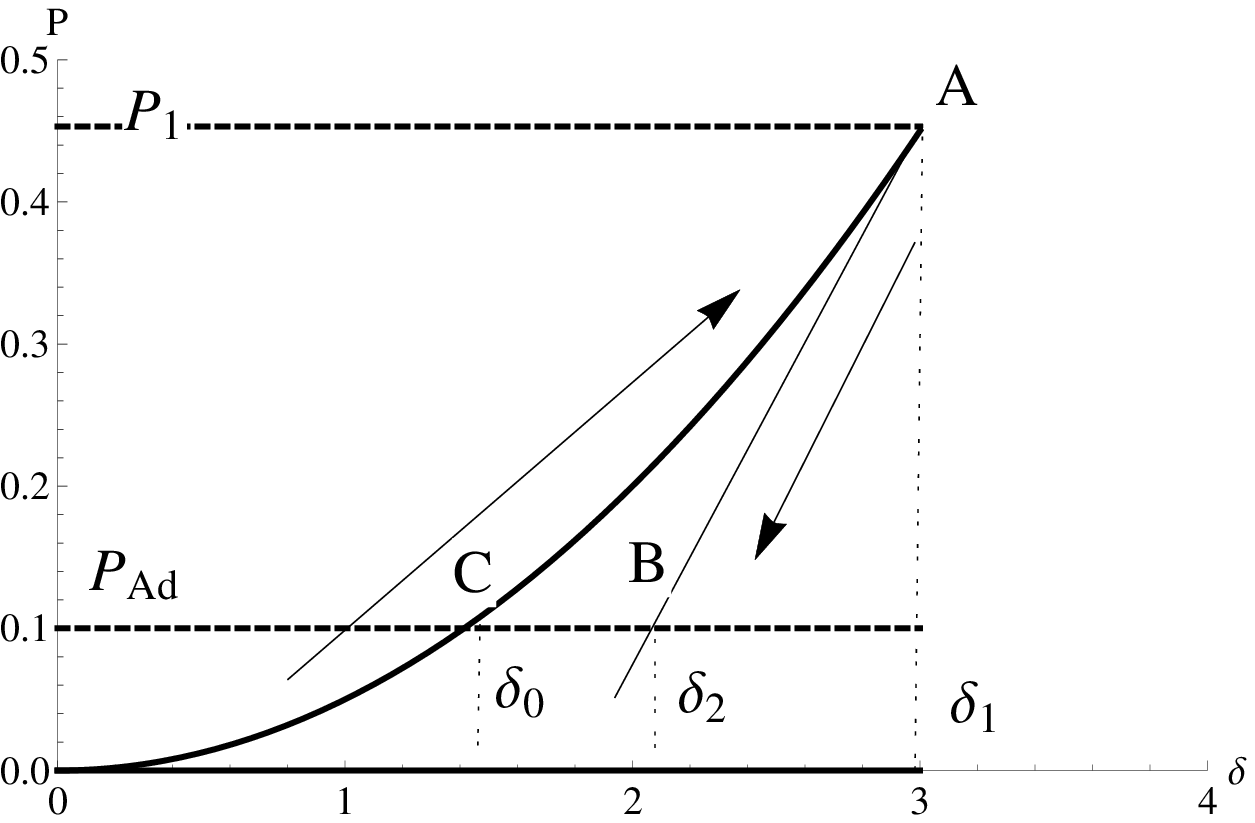}%
}
&
\end{tabular}

Fig.2. The loading scenario to obtain a general JKR solution for any contact
problem for which we know the adhesionless solution. (i) A "repulsive" loading
is executed without adhesive forces until the contact area is a given value
(load path $\overline{OA}$ as in the original JKR paper); (ii) A rigid-body
displacement is superposed at constant total contact area $A$ (load path
$\overline{AB}$ as in the original JKR paper)
\end{center}

The underlying assumption is that the contact area distributions are the same
as under adhesiveless conditions (for an appropriately increased normal load).
In general, the stress intensity factors will not be exactly equal at all
contact edges, but for special symmetry they are, and the solution is exact.
The solution states simply (see the original paper for details) that the
indentation is%
\begin{equation}
\delta=\delta_{1}-\sqrt{2wA^{\prime}/P_{1}^{\prime\prime}}\label{ciava1}%
\end{equation}
where $\delta_{1}$ is the adhesiveless indentation, $A^{\prime}$ is the first
derivative of contact area and $P_{1}^{\prime\prime}$ the second derivative of
the adhesiveless load with respect to $\delta_{1}$. The solution only requires
macroscopic quantities, and not very elaborate local distributions. The
solution is completed by finding the adhesive load as
\begin{equation}
P=P_{1}-\left(  \frac{\partial P_{1}}{\partial\delta_{1}}\right)
\sqrt{2wA^{\prime}/P_{1}^{\prime\prime}}\label{ciava2}%
\end{equation}

Hence, we would need an exact solution for a ball-in-a-socket. This is not
known but in principle one could take a FEM model and derive an approximation
as good as desired quite easily with an adhesionless code. An approximate
solution, with the advantage of being a simple closed form result, is given by
Liu et al. (2006), and another by Fang et al.(2015). Both solutions go beyond
the Hertz theory which starts to be invalid if the diametric ratio of ball
socket to sphere is less than a certain threshold, and Steuermann's theory
(1939) which assumes axisymmetric even-order polynomial with the form
$A_{n}r^{2n}$, but retains the half-space approximation. \bigskip We shall
make use of Liu's theory as it is much simpler, and sufficiently accurate for
our purposes.

Liu derives an approximate solution for small clearance  based on the
following assumptions:-

1. The sphere is equivalent to a rigid taper, and the contact stress
distribution in the $z$ direction along the profile of the rigid taper is
ellipsoidal as given by the Hertz theory.

2. The spherical cavity is modeled by a simple Winkler elastic foundation with
depth $R_{2}$ and stiffness $K$ rather than an elastic half-space, which rests
on a rigid base and is compressed by the rigid taper.

3. The shape of the contact area between equivalent rigid taper and the
elastic foundation satisfies the geometric relation given by Eq. 12 of Liu et
al. (2006), which is based essentially on writing geometrical condition of contact.

\section{The adhesive model}

\subsection{Review of Liu adhesiveless solution}

Liu's model results in a load-indentation relationship (we omit the subscript
"1" but it should be borne in mind that this is obviously the adhesionless
solution)
\begin{equation}
P=\frac{4\pi E^{\ast}R_{2}\delta^{2}}{5\left(  \Delta R+\delta\right)  }%
\frac{\sqrt{2\Delta R+\delta}}{\sqrt{2\left(  \Delta R+\delta\right)  }}%
\end{equation}
where $E^{\ast}$ is plane strain elastic modulus of the materials pair, and
which is
\begin{equation}
\frac{1}{E^{\ast}}=\frac{1-\nu_{1}^{2}}{E_{1}}+\frac{1-\nu_{2}^{2}}{E_{2}}%
\end{equation}

This equation is plotted in Fig.3, compared to the Hertz solution
$P_{Hertz}=\frac{4}{3}\frac{R}{\sqrt{\Delta R}}E^{\ast}\delta^{3/2}$ --- it is
clear that the Liu approximation should be used only to model the large
indentation, where it predicts (correctly) a tendency towards linear regime
where the contact area tends to the receding limit of the contact area, see
Ciavarella et al. (2006). Instead the low limit asymptotic Liu theory predicts
$P_{low}=\frac{4\pi E^{\ast}R_{2}\delta^{2}}{5\Delta R}$ and at large loads
$P_{high}=\frac{2\sqrt{2}}{5}\pi E^{\ast}R_{2}\delta$ which no longer depends
on the clearance. The asymptotic trends are obvious in the Fig.1a but are not
represented for clarity of comparison with Hertz.

\begin{center}%
\begin{tabular}
[c]{ll}%
{\includegraphics[
height=3.1631in,
width=5.056in
]%
{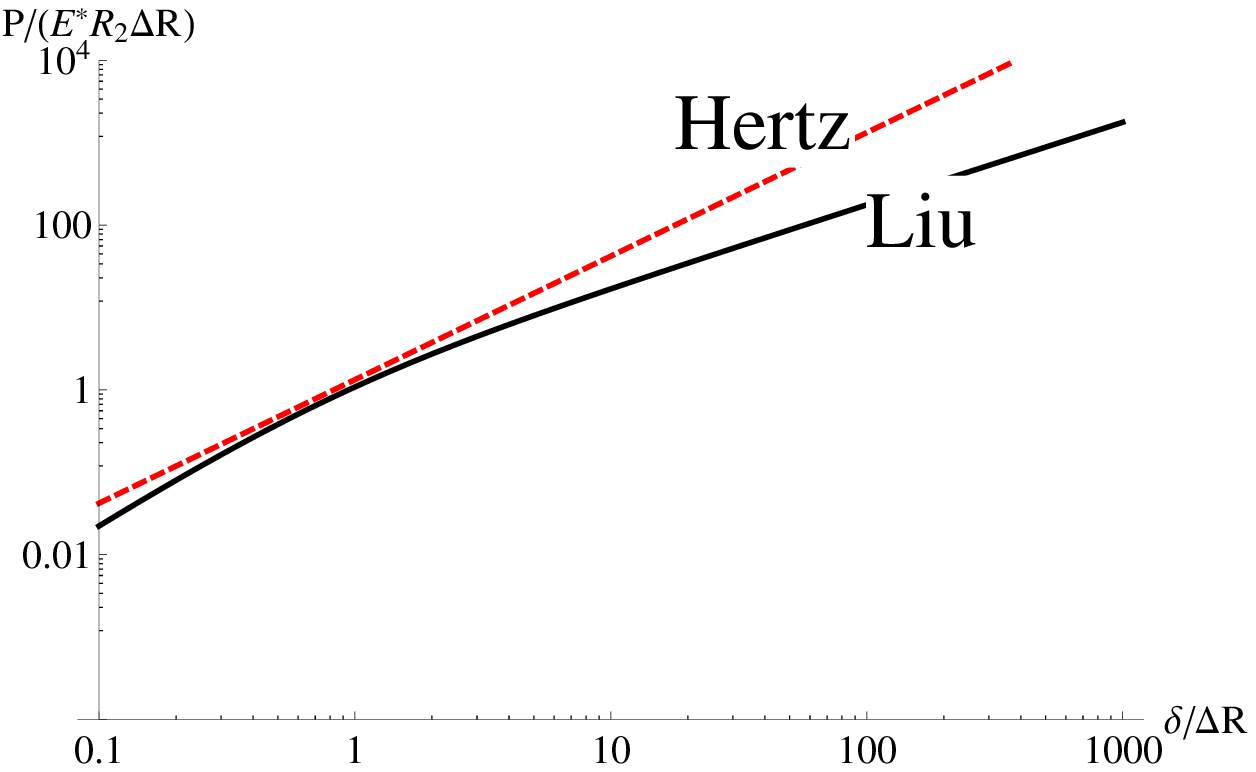}%
}
& (a)\\%
{\includegraphics[
height=3.203in,
width=5.056in
]%
{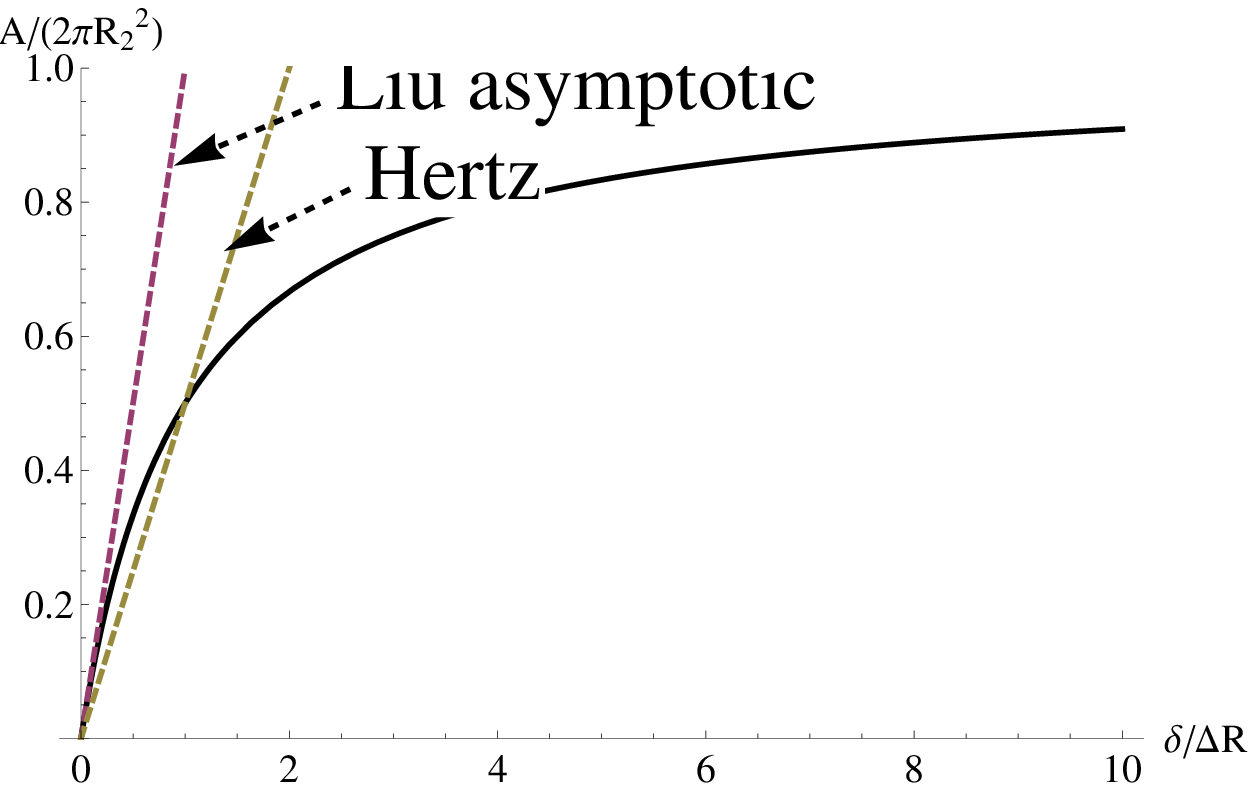}%
}
& (b)
\end{tabular}

Fig.3 - (a) load indentation relationship of Liu et al. (2006), as compared to
the Hertz classical solution which is correct at small indentations (Johnson,
1985); (b) contact area as a function of indentation for Liu's solution (also
in asymptotic form), compared to Hertz 
\end{center}

As we need the second derivative, we obtain this easily in closed form
\begin{equation}
\frac{\partial^{2}P}{\partial\delta^{2}}=\frac{\sqrt{2}}{10}\pi E\Delta
R^{2}R_{2}\frac{\left(  32\Delta R^{2}+16\Delta R\delta-\delta^{2}\right)
\allowbreak}{\left(  \delta+\Delta R\right)  ^{\frac{7}{2}}\left(
\delta+2\Delta R\right)  ^{\frac{3}{2}}}%
\end{equation}
which starts of as constant in the regime where we don't need to use Liu's
solution, and ends up to zero where the load becomes truly linear with indentation.

More complex is the relation between load parameter and semi-angle of contact
since this requires additional parametric dependences. Introducing
$\lambda=\sin\left(  \varepsilon/2\right)  $ , it can be expressed as
\begin{equation}
\frac{ER_{2}\Delta R}{P}=\frac{5\left(  1-2\lambda^{2}\right)  }{16\pi
\lambda^{4}\sqrt{1-\lambda^{2}}}%
\end{equation}
and the area of contact is obviously
\begin{equation}
A=2\pi R_{2}^{2}\left(  1-\cos\varepsilon\right)
\end{equation}
Therefore, $A=A\left(  \varepsilon\right)  $, $\arcsin\lambda=\frac
{\varepsilon}{2}$, $P=P\left(  \lambda\right)  $ so that one could obtain
$\frac{\partial A}{\partial\delta}=\frac{\partial A}{\partial\varepsilon}%
\frac{\partial\varepsilon}{\partial\lambda}\frac{\partial\lambda}{\partial
P}\frac{\partial P}{\partial\delta}$ by the chain rule. However, this results
in an expression which doesn't depend only on $\delta$ and therefore there is
no alternative to a numerical differentiation. A plot of the contact area for
Liu's solution, comparison with the Hertz one, is given in Fig.3b. Obviously,
there is again an error at low loads.

If we take for reference the asymptotic values at low loads, we can write%
\begin{align}
A\left(  \delta\right)    & =2\pi R_{2}^{2}f_{A}\left(  \frac{\delta}{\Delta
R}\right)  \label{nondim1}\\
P\left(  \delta\right)    & =E^{\ast}R_{2}\delta f_{P}\left(  \frac{\delta
}{\Delta R}\right)  \label{nondim2}%
\end{align}

\subsection{Derivation of the adhesive solution}

Using the general JKR solution (\ref{ciava1},\ref{ciava2}) we obtain (now $P$
stands for the adhesive load), using the dimensionless functions
(\ref{nondim1},\ref{nondim2})
\begin{equation}
P=E^{\ast}R_{2}\Delta Rf_{P}\left(  \frac{\delta}{\Delta R}\right)  -E^{\ast
}R_{2}f_{P}^{\prime}\left(  \frac{\delta}{\Delta R}\right)  \sqrt{2\frac
{w}{E^{\ast}}R_{2}\frac{f_{A}^{\prime}\left(  \frac{\delta}{\Delta R}\right)
}{f_{P}^{\prime\prime}\left(  \frac{\delta}{\Delta R}\right)  }}%
\end{equation}
Therefore, if we normalize by a reference pull-off value $P_{0}=2\pi
w\frac{R_{2}^{2}}{\Delta R}$ which we shall see is the asymptotic value for
this model (see Appendix), we obtain%
\begin{equation}
\frac{P}{P_{0}}=\theta^{2}f_{P}\left(  \frac{\delta}{\Delta R}\right)
-\frac{\theta}{\sqrt{2\pi}}f_{P}^{\prime}\left(  \frac{\delta}{\Delta
R}\right)  \sqrt{2\frac{f_{A}^{\prime}\left(  \frac{\delta}{\Delta R}\right)
}{f_{P}^{\prime\prime}\left(  \frac{\delta}{\Delta R}\right)  }}%
\end{equation}
where we have introduced an adhesion parameter
\begin{equation}
\theta=\frac{\Delta R}{\sqrt{2\pi wR_{2}/E^{\ast}}}%
\end{equation}

The solution depends entirely on this adhesion parameter. Fig.4 shows the
adhesive load vs contact area relationship obtained for an example case of
$\theta=1$ meant to show how the contact behaves, very similarly to the case
of a single sinusoid or a dimple geometry (Johnson, 1995, McMeeking et al., 2010).

\bigskip\ The contact can switch between a state of "weak" adhesion and one of
"strong adhesion" depending on pressure reached during loading (this is the
case when $\theta>0.82$). In particular, starting from rest (the origin in
Fig.4a), there is a first phase where the contact jumps spontaneously to point
A at zero load. From this point, we can have two scenarios: in one we start
unloading, following the curve AC, and at point C we have a pull-off jumping
into separation, very close to the prediction of the JKR theory using Hertz
geometry. The other option is if we keep loading along the branch AB, in which
case we reach a maximum compressive load which leads then to spontaneous jump
into "full contact" --- which in fact here is the condition for the area given
by the adhesiveless receding contact limit angle, which is close to 90$%
{{}^\circ}%
$ which explains the value very close to the surface of the hemisphere reached
by the contact area.

Fig.4b shows then the solution for various $\theta=0.5,1...4.5$; in
particular, the transition to "strong" adhesion regime occurs spontaneously
for $\theta<0.82$ and no "weak" adhesion regime is possible for such values.
This is a similar behaviour to that of the sinusoid or of the dimple geometry.

\begin{center}%
\begin{tabular}
[c]{ll}%
{\includegraphics[
height=2.714in,
width=5.056in
]%
{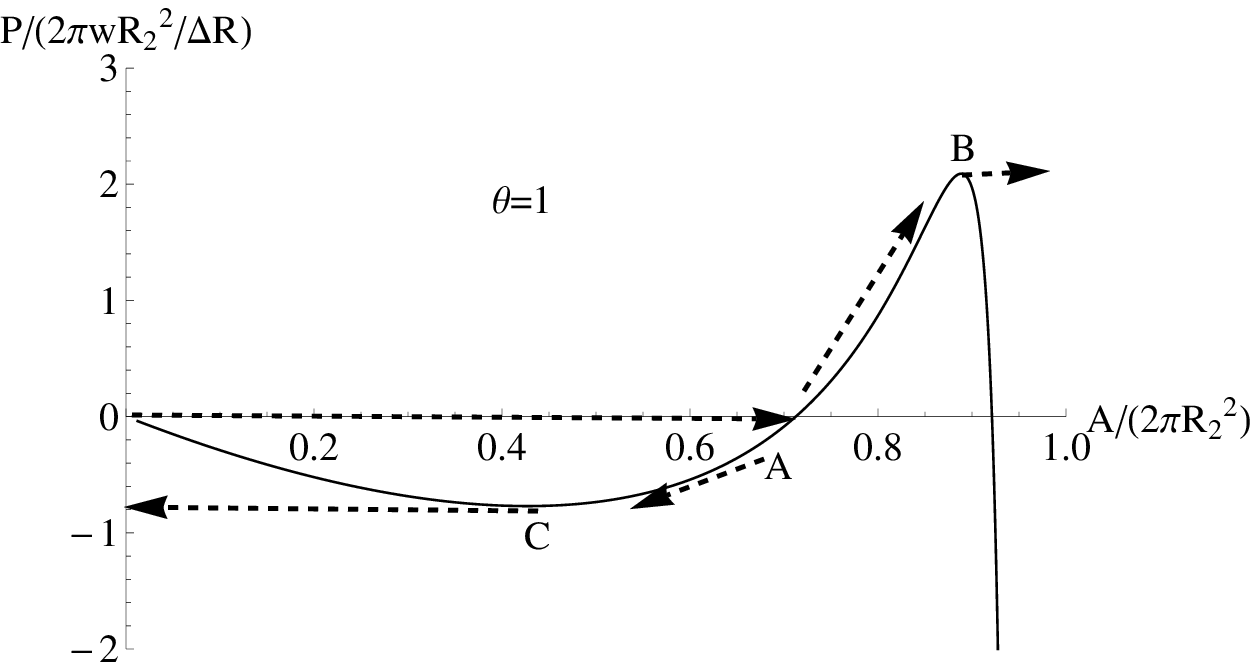}%
}
& (a)\\%
{\includegraphics[
height=2.6293in,
width=5.056in
]%
{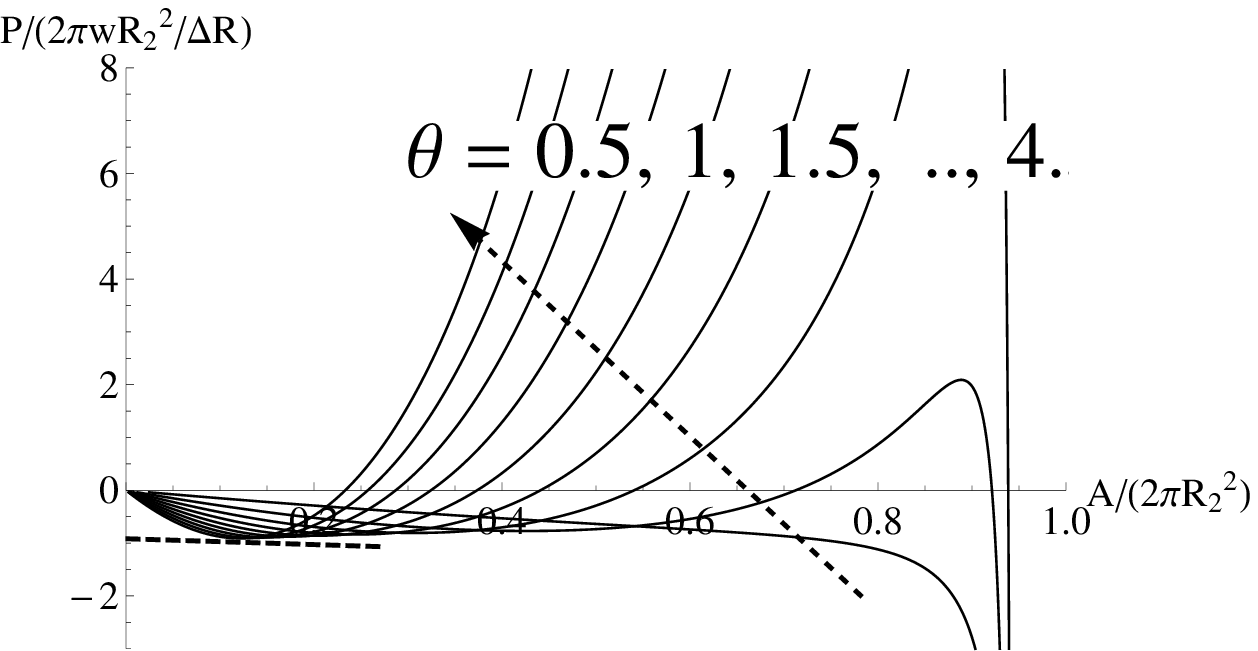}%
}
& (b)
\end{tabular}

Fig.4 - (a) Adhesive load vs contact area relationship obtained for $\theta=1$
showing that the contact can switch between a state of "weak" adhesion and one
of "strong adhesion" depending on pressure reached during loading (this is the
case when $\theta>0.82$). (b) For various $\theta=0.5,1...4.5$; transition to
"strong" adhesion regime occurs spontaneously for $\theta<0.82$ and no "weak"
adhesion regime is possible. 
\end{center}

\section{Discussion}

For a single sinusoid of wavelength and amplitude $\lambda,h$, Johnson (1995)
finds the JKR (Johnson \textit{et al.}, 1971) regime is governed by a single
parameter, $\alpha$, defined as
\begin{equation}
\alpha=\sqrt{\frac{2}{\pi^{2}}\frac{w\lambda}{E^{\ast}h^{2}}}\label{alfa-KLJ}%
\end{equation}
where $\alpha$ represents the square of the ratio of the surface energy in one
wavelength to the elastic strain energy when the wave is flattened. Solving
the problem, it turns out that for $\alpha>0.57$, there is a spontaneous snap
into full contact, and from this state, detachment should occur only at values
of stress close to theoretical strength. This "paradoxical" behaviour required
Johnson to postulate that (especially for 2D roughness), there would be a
limit due to entrapment of air in the valleys, or contaminants, or short scale
roughness. In fact, there is also the limit of the JKR assumption, and a more
general model would require a Maugis law of attractive and a cohesive solution
(for the dimple geometry, for example, Papangelo \& Ciavarella, 2017a). Also,
the JKR assumption, with the presence of roughness, not necessarily leads to
the expected reduction of stickiness and on the contrary an enhancement is
possible for special geometries of roughness (for the dimple, see Papangelo \&
Ciavarella 2017b).

In our case, an interesting interpretation is rather based on load: indeed, if
we take the Hertzian load at indentation $\delta=\Delta R$
\begin{equation}
\theta=\sqrt{\frac{\Delta R^{2}}{2\pi wR_{2}/E^{\ast}}}=\left(  3/4\right)
\sqrt{\frac{P_{Hertz}\left(  \frac{\delta}{\Delta R}=1\right)  }{P_{JKR}}}%
\end{equation}
and therefore, strong adhesion found for $\theta<0.82$ corresponds to
$P_{JKR}>0.84P_{Hertz}\left(  \frac{\delta}{\Delta R}=1\right)  $. In other
words\textit{, strong adhesion is reached for cases for which the JKR adhesive
pull-off load is of the same order as the Hertzian load to produce an
indentation equal to the clearance. }Also, notice that the load at pull-off in
the "weak" regime is marginally affected by the value of $\theta$ (and not in
a manner which is worth investigating with an approximate model anyway).

Sundaram et al. (2012)\ have discussed with great precision the corresponding
case of a JKR adhesion in cylindrical contacts, where closed form solutions
are possible. They find that contact sizes exceeding the critical (maximum)
size seen in adhesionless contacts are possible, which they call
"supercritical" regime --- although this may be more of a curiosity, because
it is unclear how these can be reached. The qualitative behaviour seems
different however in some other respects, since these jumps into full contact
do not seem to appear.

\section{Conclusion}

For the contact geometry of a ball-in-socket (a ball joint), we have provided
an exact JKR solution in terms of the adhesionless solution, which becomes
approximate however because the latter requires a numerical or approximate
solution. We have found some interesting bi-stability properties of the
contact, showing a "strong" and a "weak" regime, depending on the load and on
the value of a single adhesion parameter governing the problem, which depends
on radius of the sphere, clearance, and elastic modulus as well as surface energy.

\section{References}

Brockett, C., Williams, S., Jin, Z., Isaac, G., \& Fisher, J. (2007). Friction
of total hip replacements with different bearings and loading conditions.
Journal of Biomedical Materials Research Part B: Applied Biomaterials, 81(2), 508-515.

Ciavarella, M. (2017). An approximate JKR solution for a general contact,
including rough contacts. arXiv preprint arXiv:1712.05844.

Ciavarella, M., Baldini, A., Barber, J. R., \& Strozzi, A. (2006). Reduced
dependence on loading parameters in almost conforming contacts. International
Journal of Mechanical Sciences, 48(9), 917-925.

Fang, X., Zhang, C., Chen, X., Wang, Y., \& Tan, Y. (2015). A new universal
approximate model for conformal contact and non-conformal contact of spherical
surfaces. Acta Mechanica, 226(6), 1657-1672.

Johnson, K. L.,  (1987). Contact mechanics. Cambridge university press.
Cambridge (UK).

Johnson, K. L., K. Kendall, and A. D. Roberts. (1971). Surface energy and the
contact of elastic solids. Proc Royal Soc London A: 324. 1558.

Johnson, K. L. (1995). The adhesion of two elastic bodies with slightly wavy
surfaces. Int. J. Solids Structures, 32\textbf{\ (}No. 3/4\textbf{)}, 423-430.

Liu, C. S., Zhang, K., \& Yang, L. (2006). Normal force-displacement
relationship of spherical joints with clearances. Journal of Computational and
Nonlinear Dynamics, 1(2), 160-167.

McMeeking, R. M., Ma, L., \& Arzt, E. (2010). Bi-Stable Adhesion of a Surface
with a Dimple. Advanced Engineering Materials, 12(5), 389-397.

Papangelo, A., \& Ciavarella, M. (2017a). A Maugis--Dugdale cohesive solution
for adhesion of a surface with a dimple. Journal of The Royal Society
Interface, 14(127), 20160996.

Papangelo, A., \& Ciavarella, M. (2017b). Adhesion of surfaces with wavy
roughness and a shallow depression. Mechanics of Materials.

Steuermann, E. (1939). To Hertz's theory of local deformations in compressed
elastic bodies. In CR (Dokl.) Acad. Sci. URSS (Vol. 25, No. 5, pp. 359-361).

Sundaram, N., Farris, T. N., \& Chandrasekar, S. (2012). JKR adhesion in
cylindrical contacts. Journal of the Mechanics and Physics of Solids, 60(1), 37-54.

\bigskip

Wang, S., Wang, F., Liao, Z., Wang, Q., Liu, Y., \& Liu, W. (2015). Study on
torsional fretting wear behavior of a ball-on-socket contact configuration
simulating an artificial cervical disk. Materials Science and Engineering: C,
55, 22-33.

\section{Appendix. Asymptotic values}

If we use only the asymptotic Liu solution%
\begin{equation}
P=P_{1}+\left(  \frac{\partial P}{\partial\delta}\right)  _{\delta_{1}}\left(
\delta-\delta_{1}\right)  =\frac{4}{5}\pi E^{\ast}\frac{R_{2}}{\Delta R}%
\delta_{1}^{2}-\frac{8}{\sqrt{10}}\pi E^{\ast}\sqrt{l_{a}}\frac{R_{2}^{3/2}%
}{\Delta R}\delta_{1}%
\end{equation}

To find the minimum%
\[
\frac{\partial P_{Ad}}{\partial\delta_{1}}=\frac{8}{5}\pi E^{\ast}\frac{R_{2}%
}{\Delta R}\delta_{1}-\frac{8}{\sqrt{10}}\pi E^{\ast}\sqrt{l_{a}}\frac
{R_{2}^{3/2}}{\Delta R}=0
\]
which results in
\[
\delta_{1po}=\frac{5}{\sqrt{10}}\sqrt{l_{a}}R_{2}^{1/2}%
\]
so the load is
\begin{equation}
P_{Ad,po}=-2\pi w\frac{R_{2}^{2}}{\Delta R}=-2\pi wR_{eq}%
\end{equation}
where of course we recognize that the equivalent radius of Hertz' theory
\[
\frac{1}{R_{eq}}=\frac{1}{R_{1}}-\frac{1}{R_{2}}=\frac{R_{2}-R_{1}}{R_{1}%
R_{2}}\simeq\frac{\Delta R}{R_{2}^{2}}%
\]

In other words, the Liu model leads to a DMT result rather than JKR, with the
difference being due to the fact that it doesn't correspond to the Hertz
theory in the limit of large clearance, as evident from Fig. 9 of their paper. 

\bigskip

\end{document}